\documentstyle[floats,aps]{revtex}

\begin{document}
\draft

\twocolumn[\hsize\textwidth\columnwidth\hsize\csname@twocolumnfalse\endcsname
\title{Quantum decoherence effect and neutrino oscillation}
\author{{\sc C.P.Sun}$ ^{1,2}$ and  {\sc D.L.Zhou}}
\address{Institute of Theoretical Physics, Academia Sinica, Beijing 100080,
P. R. China}

\maketitle
\begin{abstract}

In the view of the quantum dynamic theory of measurement developed from the
Hepp-Colemen (HC) model  (K. Hepp, Hev.Phys.Acta, {\bf 45}, 237 (1972)),
the quantum decoherence in neutrino flavor oscillation
caused by an environment surrounding neutrinos is generally considered in 
this paper. The Ellis, Hagelin, Nanopoulos and Srednicki (EHNS)
mechanism for solving the solar neutrino problem can be comprehended in a
framework of the ordinary quantum mechanics. In the weak- coupling limit, a
microscopic model is proposed to describe the transition of two neutrino
system from a pure state to a mixed state. It gives the modified formula of
survival probability of neutrino oscillation with two additional time-dependent 
parameters. For specified environments, this result shows that the oscillating 
phenomena of neutrino still exist even without a mass difference in free neutrino.

\end{abstract}

]

It is well-known that environment may decohere a quantum system immersed in
it, making this system to lose its coherence and to transit from a pure
state to a mixed state [ 1]. Based on the quantum dynamic approach of this
environment influences [2], which was developed from the Hepp-Colemen (HC)
theory of quantum measurement and its clarified physical representation [3],
the decoherence phenomenon is investigated for the neutrino flavor
oscillation [4] in this paper.

Neutrino oscillation is a subject of much intense theoretical and
experimental research. In comparison with the prediction from the standard
solar model(SSM), the deficit of solar neutrinos observed in the earth can
probably be explained by neutrino oscillations with a (mass)$^2$ difference $%
\Delta m^2\sim 10^{-6}{\rm eV}^2$ [4]. Two possible mechanisms were further
\ presented to enhance the vacuum neutrino oscillations. As a matter
enhancement effect of oscillations, the Mikheyev-Smirnov-Wolfenstein (MSW)\
mechanism [ 5], was invoked when the neutrinos pass through the sun
adiabatically and non-adiabatically [6]. Initialed by Ellis, Hagelin,
Nanopoulos and Srednick (EHNS) in 1983 [7], there was another mechanism (
now called EHNS\ mechanism) to modify neutrino oscillations. Similar to the
quantum decoherence in generic problem of environment influence[1-3], the
EHNS\ mechanism makes a transition of neutrino from a pure state to a mixed
state due to some unknown couplings of neutrinos to the environment
surrounding them. Such an environment may be the background field caused by
Hawking$^{^{\prime }}$s quantum gravity effect, like the evaporation of
black holes [8]. With irreversible elements characterized by
phenomenological parameters, the original EHNS\ mechanism suggested an
effective von Neumann equation of density matrix. Its solutions modify the
survival probability of neutrino oscillation . Following this work some
significant efforts were made by different authors in preserving the
linearity, locality, conservation of probability and unitarity [9]. Notice
that the EHNS parameters were determined to some extent either by analyzing
the experimental data of CP violation in $K^0-\overline{K^0}$ system [10] or
based on further theoretical considerations, such as the string theory with
an Einstein-Yang-Mills black hole in four dimensions[11].

These investigations reflect the successful aspects of the EHNS\ mechanism
in solving the solar neutrino problem. However, as the phenomenological
motion equation of density matrix concerns quantum irreversible process, it
apparently violates quantum mechanics since the Schroedinger equation is
reversible with time-reversal symmetry. As a correct physical theory to
solve the solar neutrino problem, the EHNS mechanism has to be consistent
with quantum mechanics. To remove this confliction in this paper we show
that rationality of the EHNS mechanism is related to the reduction of
quantum mechanics for the ``universe'' , a total system formed by the
neutrino system plus environment. The physical essence of our observation is
that the couplings of neutrino to the EHNS\ mechanism results in quantum
decoherence of the flavor eigenstates of neutrino, which are coherent
superpositions of neutrino mass eigenstates if there is not an external
influence. Mathematically, this physical process is described by a reduced
density matrixes of neutrino system by tracing out over the variables of the
environment. The calculation of the reduced density matrixes explicitly
leads to a modified formula of survival probability of neutrino oscillation.
It not only introduces extra parameters to neutrino oscillation, but also
implies a novel dynamic effect that the oscillating phenomena of neutrino
may exists even without a mass difference in free neutrino.

Methodologically, the starting point of the present study is the quantum
dynamic approach [2] for decoherence problem(or wave function collapse, WFC)
in quantum measurement. We recall that, in the traditional theory of quantum
measurement , the WFC postulate is only an extra postulate added to the
ordinary quantum mechanics. Under this postulate, once we measure an
observable and obtain a definite value $a_k$ the state of system must
collapse into the corresponding eigen-state $|k\rangle $ from a coherent
superposition $|\phi \rangle =\sum_kc_k|k\rangle \langle k|$. Through
density matrix this process is described by a quantum decoherence process $%
\rho =|\phi \rangle \langle \phi |\rightarrow $ $\hat{\rho}%
=\sum_k|c_k|^2|k\rangle \langle k|$ from a pure state to a mixed state. In
quantum dynamic approach, both the measured system and the measuring
apparatus obey Schroedinger equation and the dynamic evolution governed by
their interaction is supposed to result in quantum decoherence for the
reduced dynamics of the measured system under certain conditions, such as
the macroscopic limit that the measuring apparatus contains an infinite
number of subsystems. Notice that the environment coupling to neutrino
usually can be considered as a bath of infinite degrees of freedom. For
quantum decoherence, there is a strong resemblance between the neutrino
system plus the environment and the measured system plus the measuring
apparatus. So we can study quantum decoherence effect in neutrino
oscillation by making use of the quantum dynamic approach developed in
treating quantum measurement. In the following our discussion will focus on
the case of two generations, but it can be easily generalized to the case of
three generations.

The neutrino weak interaction (flavor) eigenstates $\nu _e$ and $\nu _\mu $
do not coincide with their mass eigenstates $\nu _{1\text{ }}$and $\nu _2$
with masses $m_1$ and $m_2$. Through a vacuum mixing angle $\theta $, the
neutrinos mixture is described as a two dimensional rotation 
\begin{equation}
\nu _e=\nu _1\cos \theta +\nu _2\sin \theta
\end{equation}
\[
\nu _\mu =-\nu _1\sin \theta +\nu _2\cos \theta \quad 
\]
Without the interaction between neutrinos and other objects, the vacuum
neutrino oscillation result from free Hamiltonian 
\begin{equation}
H_n=E_1\bar{\nu}_1\nu _1+E_2\bar{\nu}_2\nu _2  \label{2}
\end{equation}
As a result, the survival probability that an electron-neutrino remain as
itself is: 
\begin{equation}
p_{\nu _e\rightarrow \nu _e}(t)\simeq 1-\sin ^2(2\theta )\sin ^2[\frac{%
\delta m^2t}{4E}]  \label{3}
\end{equation}
Here $\delta E=E_2-E_1,$ $\delta m^2=m_2^2-m_1^2$ and we have considered
masses of neutrinos are so small that their energy and momentum are very
close. The above well-known result was obtained completely from the coherent
superposition Eq.(1) of pure states $\nu _{1\text{ }}$and $\nu _2$ and it
will be certainly modified if the mixed states of neutrinos were introduced
from fluctuating interactions.

Now, we consider the coupling of neutrinos to environment, which may be a
background provided by the certain gravity effect such as the Hawking's
evaporation of black holes. In general , without referring to the concrete
forms of the environment Hamiltonian $H_G(q)$ of dynamic variables $q$ , we
can roughly write down the interaction Hamiltonian in the weakly -coupling
limit

\begin{equation}
H_I=[\lambda _1\bar{\nu}_1\nu _1+\lambda _2\bar{\nu}_2\nu _2]q  \label{4}
\end{equation}
which is linear for both the density projections $\bar{\nu}_i\nu _i(i=1,2)$
of neutrinos and the environment variable $q$. The coupling constants $%
\lambda _i$ $(i=1.2)$satisfying $\lambda _1\neq \lambda _2$ implies  that
the couplings to environment have different strengths for different
neutrinos $\nu _1$ and $\nu _2$. For instance, in the quantum gravity
background, perhaps there is $\lambda _i\propto m_i^2.$ Up to the first
order perturbation for quantum dissipation problem , Caldeira and Leggett
[12] have pointed out that, any environment weakly coupling to a system may
be approximated by a bath of oscillators with linear interaction in the
weakly coupling condition that ``each environmental degree of freedom is
only weakly perturbed by its interaction with the system''. For both quantum
decoherence [2,14]as well as quantum dissipation [12,13] , we have modeled a
general environment with the linear coupling Hamiltonians similar to Eq.(4)
and discussed the universality of such modeling scheme.

Because the total system, governed by the total Hamiltonian $H=H_n+H_G+H_I$
, is closed, the quantum mechanics can work well to describe its dynamic
evolution. Suppose the neutrino is initially in a flavor state $\nu _e$ with
a pure density matrix 
\begin{equation}
\rho _{\nu _e}=\bar{\nu}_e\nu _e=\left( 
\begin{array}{cc}
\cos ^2\theta  & \cos \theta \sin \theta  \\ 
\cos \theta \sin \theta  & \sin ^2\theta 
\end{array}
\right)   \label{5}
\end{equation}
and the environment in a general initial state of density matrix $\rho _G,$%
the initial condition for the total system should be : 
\begin{equation}
\rho (t=0)=\rho _{\nu _e}\otimes \rho _G  \label{6}
\end{equation}
By considering the quantum non-demolition features of the interaction ,
i.e., $[H_I,H_n]=0,$ $[H_I,H_G]\neq 0,$the total density matrix can be
directly computed as

\[
\rho (t)=\cos ^2\theta \cdot \bar{\nu}_1\nu _1\otimes U_1(t)\rho
_GU_1^{\dagger }(t)+ 
\]
\begin{equation}
+\sin ^2\theta \cdot \bar{\nu}_2\nu _2\otimes U_2(t)\rho _GU_2^{\dagger }(t)]
\label{7}
\end{equation}
\[
+\cos \theta \sin \theta \cdot \bar{\nu}_1\nu _2\otimes U_1(t)\rho
_GU_2^{\dagger }(t)\exp (i\delta Et)+H.c. 
\]
where 
\begin{equation}
U_i(t)=\exp \{-i[H_G(q)+\lambda _iq]t\}\quad (i=1,2)  \label{8}
\end{equation}
for $i=1,2$ are the effective evolution operators of environment. They show
that the back-actions on environment exerted by $\nu _1$ and $\nu _2$
respectively are different for different coupling strengths $\lambda
_i(i=1,2)$. Namely, the environment can distinguishes different mass
eigenstates of neutrino.

As our main interest is only in the neutrino oscillation rather than the
motion of the environment, we trace out over the variable of environment to
obtain the reduced density matrix of the neutrino system. It is

\[
\rho _n(t)=\cos ^2\theta \cdot \bar{\nu}_1\nu _1++\sin ^2\theta \cdot \bar{%
\nu}_2\nu _2
\]
\begin{equation}
+F(t)\cos \theta \sin \theta \cdot \bar{\nu}_1\nu _2\exp (i\frac{\delta m^2t%
}{2E})+H.c.  \label{9}
\end{equation}
of the neutrino system. Here, the decoherence factor 
\begin{equation}
F(t)=Tr_g[U_2^{\dagger }(t)U_1(t)\rho _G]  \label{10}
\end{equation}
characterizes the extent that the neutrino system loses its coherence. The
reduced density matrix $\rho _n(t)$ determines the survival probability that
a electron-neutrino remain in itself is 
\[
p_{\nu _e\rightarrow \nu _e}(t)=Tr_n[\bar{\nu}_e\nu _e\rho _n(t)]
\]
\begin{equation}
=1-\frac 12\sin ^2(2\theta )\{1-\cos [\delta Et+\alpha (t)]|F(t)|\}
\label{11}
\end{equation}
where the real number $\alpha (t)$ is the phase of $F(t).$ It is easy to
prove that $|F(t)|$ is less than 1 since $Tr[\rho _G]=1$ and $U_i(t)$
(i=1,2) are unitary. In the above concise formula (11) for neutrino
oscillation, the influences of environment are simply summed up to a
decoherence factor $F(t)$, which is time-dependent while energy-independent.
If the difference $\lambda _1-\lambda _2$ in couplings of $\nu _1$ and $\nu
_2$ to environment does not depend on $\delta m$, the difference  in their
masses, the result (11) directly shows a novel fact that, even neutrino is
massless or electron- and muon- neutrinos have a same mass, it is possible
to see the oscillational behavior between electron- and muon- neutrino.
Here, the neutrino oscillation is caused by the phase $\alpha (t)$ of the
decoherence factor, that is 
\begin{equation}
p_{\nu _e\rightarrow \nu _e}(t)=1-\frac 12\sin ^2(2\theta )(1-\cos [\alpha
(t)]|F(t)|)  \label{12}
\end{equation}
As the above derived temporal phenomenon is independent of the energy of
neutrino flux, it can not solve the present solar neutrino problem with
energy -dependence spectrum .

To get one step further we consider two extreme cases. When $|F(t)|=1,$ or $%
F(t)=\exp [i\alpha (t)]$ for real $\alpha ,$ the system still in a pure
state 
\[
\rho _n(t)=\bar{\phi}(t)\phi (t):
\]
\begin{equation}
\phi (t)=\nu _1\exp (iE_1t-i\frac \alpha 2)\cos \theta +\nu _2\exp (iE_2t+i%
\frac{\alpha (t)}2)\sin \theta   \label{13}
\end{equation}
In this case the influences of environment on neutrino oscillation only
result in a time-dependent phase shift $\frac{\alpha (t)}2,i.e.,$ 
\begin{equation}
p_{\nu _e\rightarrow \nu _e}(t)=1-\sin ^2(2\theta )\sin ^2[\frac{\delta m^2t%
}{4E}+\frac{\alpha (t)}2]  \label{14}
\end{equation}
As a signature of the environment influence, this phase shift only affects
the spectrum distribution $P(E)=p_{\nu _e\rightarrow \nu _e}(t)$ in the
higher energy region. For an extreme case with $F(t)=0,$ the coherence in
neutrino system is completely lost and  the neutrinos transit into a
completely mixed state 
\begin{equation}
\rho _n(t)=\cos ^2\theta \cdot \bar{\nu}_1\nu _1++\sin ^2\theta \cdot \bar{%
\nu}_2\nu _2  \label{15}
\end{equation}
with vanishing off-diagonal elements. It leads to a constant transition
probability$\frac 12\sin ^2(2\theta )$ from $\nu _e$ to $\nu _\mu $ and the
corresponding survival probability that $\nu _e$ remain in itself is $1-%
\frac 12\sin ^2(2\theta ).$ Notice that these results only depend on the
vacuum mixing angle $\theta $ and do not fit the neutrino spectrum varying
with the energy of neutrino.

To get the idea of explicit form of $F(t)$ we illustrate the determination
of decoherence factor in the harmonic oscillator model of environment
[3,14]. Let $a_j^{+}$and $a_j$ be the creation and annihilation operators
for the j$^{,}$th harmonic oscillator in environment. The Hamiltonian of
environment takes the form $H=\sum_{j=1}^N\hbar \omega _ja_j^{+}a$ and its
interaction with neutrino can be modeled as a linear coupling: 
\begin{equation}
H_I=[\lambda _1\bar{\nu}_1\nu _1+\lambda _2\bar{\nu}_2\nu
_2]\sum_{j=1}^N\hbar g_j(a_j^{+}+a_j)  \label{16}
\end{equation}
Let the environment be initially in the vacuum state $|0\rangle
_e=|0_1\rangle \otimes |0_2\rangle \otimes \cdot \cdot \cdot \otimes
|0_N\rangle $ where $|0_j\rangle $is the ground state of the j$^{,}$th
single harmonic oscillator, the corresponding decoherence factor 
\[
F(t)\equiv F(N,t)=\prod_{j=1}^N\ \langle 0_j|U_2^{^j\dagger
}(t)U_1^j(t)|0_j\rangle \equiv \prod_{j=1}^NF_j(t)
\]
can be obtained by solving the Schroedinger equations of $U_i^j(t)$ ($i=1,2$%
) governed by the Hamiltonians of forced harmonic oscillators $H_i^j=\hbar
\omega _ja_j^{+}a_j+\lambda _ig_j(a_j^{+}+a_j).$ The result is 
\[
|F(t)|=\exp [-\sum_{j=1}^N\frac{2g_j^2(\delta \lambda )^2}{\omega _j{}^2}%
\sin ^2(\frac{\omega _jt}2)]
\]
\begin{equation}
\alpha (t)=\sum_{j=1}^N\frac{g_j^2\delta \lambda ^2}{\omega _j{}}[t+\frac{%
\sin (\omega _jt)}{\omega _j}]  \label{17}
\end{equation}
where $\delta \lambda =\lambda _2-\lambda _1,\delta \lambda ^2=\lambda
_2^2-\lambda _1^2.$The decoherence time is decided by the norm part of $%
F(N,t),$ which is the same as that from the two level subsystems model of
environment in the weakly coupling limit [2]. Especially, when the
environment consists of identical particles, we have $\omega _j=\omega $ and 
$g_j=g.$ In the van Hove limit of weakly-coupling , i.e., $g\rightarrow
0,N\rightarrow \infty $ , but $g\sqrt{N}\rightarrow $a finite constant $G,$a
simple result for decoherence factor is obtained as $|F(t)|=\exp [-(\delta
\lambda )^2\frac{2G^2}{\omega {}^2}\sin ^2(\frac{\omega t}2)]$ and $\alpha
(t)=\delta \lambda ^2[\frac{G^2}{\omega {}^2}t+\frac{\sin (\omega t)}\omega
].$ Then, an explicit formula of environment-modifying neutrino oscillation
is obtained as 
\[
p_{\nu _e\rightarrow \nu _e}(t)=1-\frac 12\sin ^2(2\theta )\times 
\]
\[
\{1-\cos [(\frac{\delta m^2}{4E}+\frac{G^2\delta \lambda ^2}{2\omega {}^2}%
)t+\delta \lambda ^2\frac{\sin (\omega t)}{2\omega }]\times 
\]
\begin{equation}
\exp [-(\delta \lambda )^2\frac{2G^2}{\omega {}^2}\sin ^2(\frac{\omega t}2%
)]\}  \label{18}
\end{equation}
In the low-frequency case of environment ($\omega \rightarrow 0),$denote  $%
\beta =\frac 12(\delta \lambda )^2G^2,$ we have an exponential decay $%
|F(t)|=\exp [-\beta t]$ and $\alpha (t)=\frac \beta {\omega {}^2}t\cdot $ So 
\[
p_{\nu _e\rightarrow \nu _e}(t)\rightarrow 1-\frac 12\sin ^2(2\theta )\times 
\]
\begin{equation}
\{1-\cos [(\frac{\delta m^2}{4E}+\frac \beta {\omega {}^2})t]\}\times \exp
[-\beta t].  \label{19}
\end{equation}

In conclusion, we have considered the quantum decoherence problem in
neutrino flavor oscillation based on our dynamic approach for quantum
measurement. The phenomenon of transition of neutrinos from a pure state to
a mixed state, similar to EHNS mechanism, has been understood to some extent
in the view of the ordinary quantum mechanics. Our study leads to the
modified neutrino oscillations with two additional time-dependent
parameters. For specified environments, they show that the oscillating
phenomena of neutrino still exist even without a mass difference in free
neutrinos. However, we have not considered yet the influences of quantum
decoherence on the MSW mechanism and the quantum dissipation effects due to
the couplings to environment in off-diagonal form $[g\bar{\nu}_1\nu _2+H.c]q$
[12, 13]. These effects perhaps are crucial to the final solution to solar
neutrino problem, and the method used by this paper can be generalized to
handle them without any difficult in principle. The main problem in present
is that the explicit calculation of the decoherence factor $F(t)$ depends
the details of the environment and its couplings to neutrinos in certain
extent, but up to date we have not a complete knowledge yet about a complex
environment. If the environment is provided by the Hawking$^{,}$s
evaporation of black holes in gravity field, there is not yet an universally
acknowledged scheme to quantize gravity. Regardless of this problem, the
present paper at least suggests a general rule to microscopically analyze
the decoherence modification of neutrino oscillation. Once the Hamiltonians
of the environment and its coupling to neutrinos, even in the
weakly-coupling case, were provided the decoherence factor can be explicitly
calculated to fit the datum in the solar neutrino problem.

\vskip 0.3cm

\noindent Acknowledgment: The authors wishe to thanks T.H. Ho, X.Q.Li,
Y.L.Wu and C.H.Chang for many helpful discussions. This work .is supported
in part by the NFS China \newline

$^1$electronic address: suncp@itp.ac.cn;

$^2$internet www site: http:// www.itp.ac.cn/\symbol{126}suncp

\end{document}